\documentclass[a4paper]{jpconf}
\usepackage{graphicx}
\usepackage{aas_macros}
\usepackage{amssymb}

\newcommand{\Rin}{R_{\rm in}}

\newcommand{\Lhalf}{L^{1/2}}
\newcommand{\Rhalf}{R_{1/2}}

\usepackage[square,sort&compress]{natbib}
\begin{document}
\title{Probing the innermost dusty structure in AGN with mid-IR
  and near-IR interferometers}


\author{M.~Kishimoto$^1$, S.~F.~H\"onig$^2$, R.~Antonucci$^2$,
  R.~Barvainis$^3$, T.~Kotani$^4$, F.~Millour$^5$,
  K.~R.~W.~Tristram$^1$, G.~Weigelt$^1$}

\address{$^1$ Max-Planck-Institut f\"ur Radioastronomie, Auf dem H\"ugel
  69, 53121 Bonn, Germany\\
  $^2$ Physics Department, University of California, 
  Santa Barbara, CA 93106, USA.\\ $^3$ National
  Science Foundation, 4301 Wilson Boulevard, Arlington, 
  VA 22230, USA.\\  
  $^4$ ISAS, JAXA, 3-1-1 Yoshinodai, Sagamihara, Kanagawa, 
  229-8510 Japan.\\
  $^5$ Observatoire de la C\^ote d Azur, 
           Departement FIZEAU, Boulevard de l'Observatoire, BP 4229, 06304, Nice
 Cedex 4, France
}

\ead{mk@mpifr-bonn.mpg.de}

\begin{abstract}
  With mid-IR and near-IR long-baseline interferometers, we are now
  mapping the radial distribution of the dusty accreting material in
  AGNs at sub-pc scales.  We currently focus on Type 1 AGNs, where the
  innermost region is unobscured and its intrinsic structure can be
  studied directly.  As a first systematic study of Type 1s, we
  obtained mid-/near-IR data for small samples over $\sim$3--4 orders
  of magnitudes in UV luminosity $L$ of the central engine.  Here we
  effectively trace the structure by observing dust grains that are
  radiatively heated by the central engine. Consistent with a naive
  expectation for such dust grains, the dust sublimation radius $\Rin$
  is in fact empirically known to be scaling with $\Lhalf$ from the
  near-IR reverberation measurements, and this is also supported by
  our near-IR interferometry. Utilizing this empirical relationship,
  we normalize the radial extent by $\Rin$ and eliminate the simple
  $\Lhalf$ scaling for a direct comparison over the samples.  We then
  find that, in the mid-IR, the overall size in units of $\Rin$ seems
  to become more compact in higher luminosity sources. More
  specifically, the mid-IR brightness distribution is rather well
  described by a power-law, and this power-law becomes steeper in
  higher luminosity objects.  The near-IR flux does not seem to be a
  simple inward extrapolation of the mid-IR power-law component toward
  shorter wavelengths, but it rather comes from a little distinct
  brightness concentration at the inner rim region of the dust
  distribution. Its structure is not well constrained yet, but there
  is tentative evidence that this inner near-IR-emitting structure has
  a steeper radial distribution in jet-launching objects. All these
  should be scrutinized with further observations.
\end{abstract}

\section{Introduction}

Using long-baseline interferometers in both the mid-IR and near-IR, we
are probing the innermost dusty structure in the putative torus in
active galactic nuclei (AGNs) at unprecedented spatial resolutions in
the IR. One of the first goals of our study has been to directly map
the {\it radial distribution} of the accreting material. We currently
focus on Type 1 AGNs, where our line of sight is thought to be more or
less face-on and the innermost region can directly be studied without
significant effects from inclination and obscuration.  We aim to
summarize below the status of our systematic Type 1 study in both the
mid-IR and near-IR.

\section{Removing the simple $L^{1/2}$ dependency}

Let us start with addressing a simple, but naive, luminosity scaling
that should be carefully considered in our study.  The mid- and
near-IR radiation in AGNs is generally thought to be the thermal
emission from dust grains that are radiatively heated by the
UV/optical emission of the central engine. Effectively, we are using
these dust grains to trace the inner structure.  When the temperature
of the observed dust grains is simply determined by the central
engine's {\it direct} illumination and the dust thermal re-emission,
the radial distance of the dust grains at a given temperature will be
proportional to $\Lhalf$.  Here we assume that the incident spectral
shape does not change with the UV luminosity $L$, which we discuss
more below.  Then our observations at a fixed wavelength would find
the size of the structure to also scale with $\Lhalf$, since they
would be sensing the dust grains of the same temperature range.

This simple $\Lhalf$ scaling may not realize due to various quantities
and properties potentially changing with luminosity.  However, at
least for the innermost dust sublimation region, this seems to hold --
the dust sublimation radius $\Rin$ as probed by the near-IR
reverberation measurements is known to be proportional to $\Lhalf$
\cite{Suganuma06}.  Here the relation is found for optical luminosity,
rather than the more relevant UV luminosity, implying the same
UV--optical spectral shape over the range of luminosities at least
approximately.  Therefore, we assume a generic spectral shape and
define the UV luminosity $L$ as a scaled optical luminosity.  This
innermost $\Lhalf$ scaling is in fact supported also by the results
from our near-IR interferometry, although there is a distinction
between our interferometric size measurements and the reverberation
measurements, as we discuss in Sect.\ref{sect-nIR}.  Based on this
empirical relationship, which matches the naive expectation, it would
be physically insightful if we normalize the radial extent in our
study by the dust sublimation radius $\Rin$ given by the $\Lhalf$ fit
to the reverberation measurements.  This will remove the simple
$\Lhalf$ scaling and let us investigate what is left behind it.

Each interferometric measurement gives a visibility at a certain
spatial frequency, or its reciprocal, spatial wavelength, which
roughly corresponds to the spatial scale probed by that particular
configuration of the interferometer (see more in
Sect.\ref{sect-mIR}). We normalize this probed spatial scale by $\Rin$
for each observed object (in this case, spatial frequency will have
the units of cycles per $\Rin$). Differences in distance, as well as
in luminosity, will be folded into the different angular sizes of
$\Rin$, where visibility curves as a function of the spatial
wavelength in units of $\Rin$ for different objects can be compared
uniformly \cite{Kishimoto09}.  The normalization gives a physical
meaning to the probed spatial scale, which makes the interferometric
observables more interpretable physically and directly.  This is quite
important since our $uv$ converage is normally poor and we still have
to deal with direct observables in Fourier space (i.e. visibility),
rather than quantities converted to those in image space.

We note that the dust sublimation radii from the reverberation
measurements are systematically smaller by a factor of $\sim$3
\cite{Kishimoto07} than that expected for a representative property of
dust grains in the Galactic interstellar medium for each given $L$
\cite{Barvainis87}. We have specifically discussed this issue in
\cite{Kishimoto07}, including the case for large grains as well as
anisotropic radiation and absorption. The case for large grains in the
nuclear region has long been suggested on different grounds
(e.g. \cite{Maiolino01II,Gaskell04}), while the effect of anisotropy
is extensively studied recently by \cite{Kawaguchi10}.


\begin{figure}
\centering
(a)\hspace{-0.6cm}%
\includegraphics[width=0.5\textwidth]{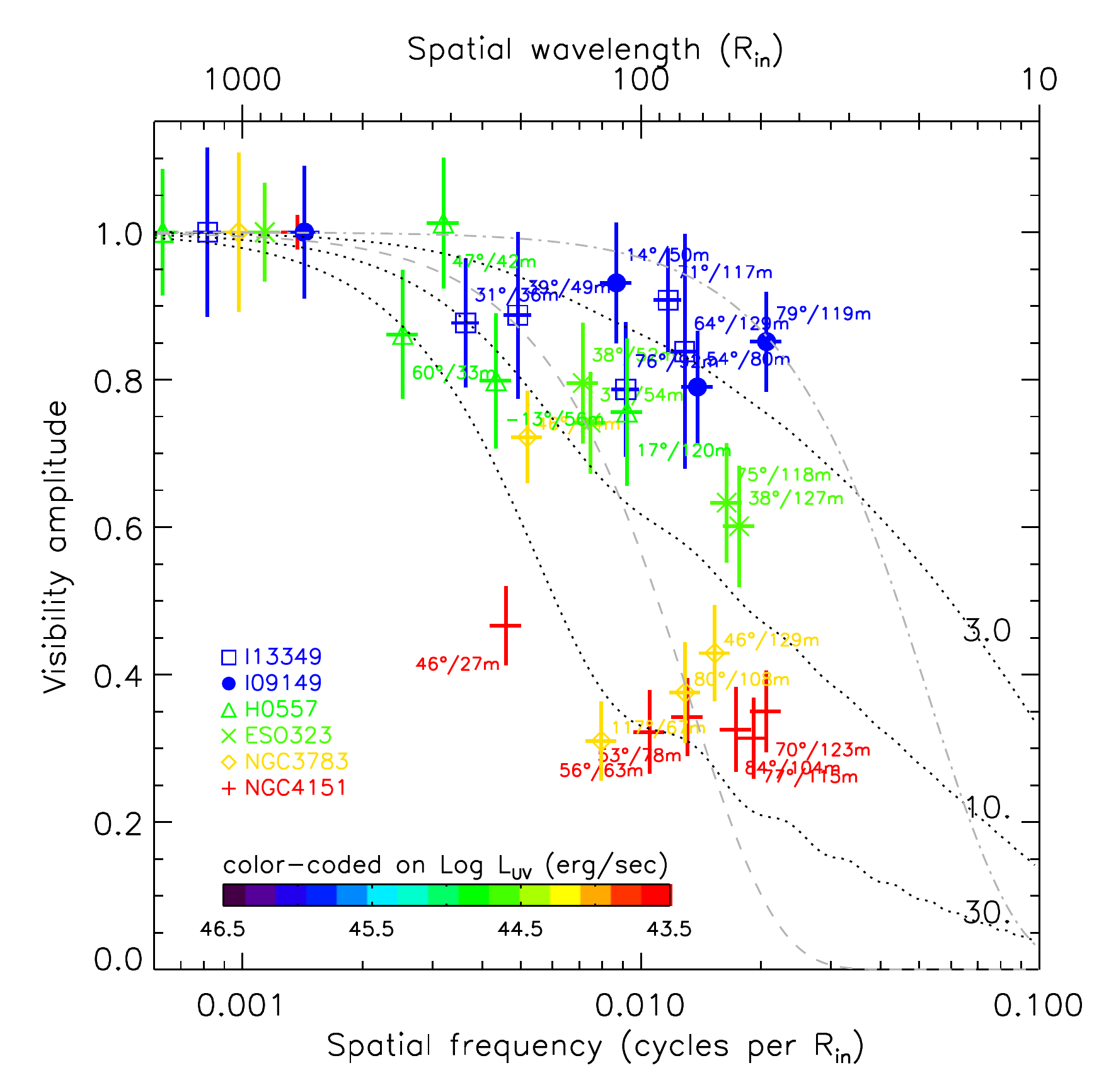}%
(b)\hspace{-0.6cm}%
\includegraphics[width=0.5\textwidth]{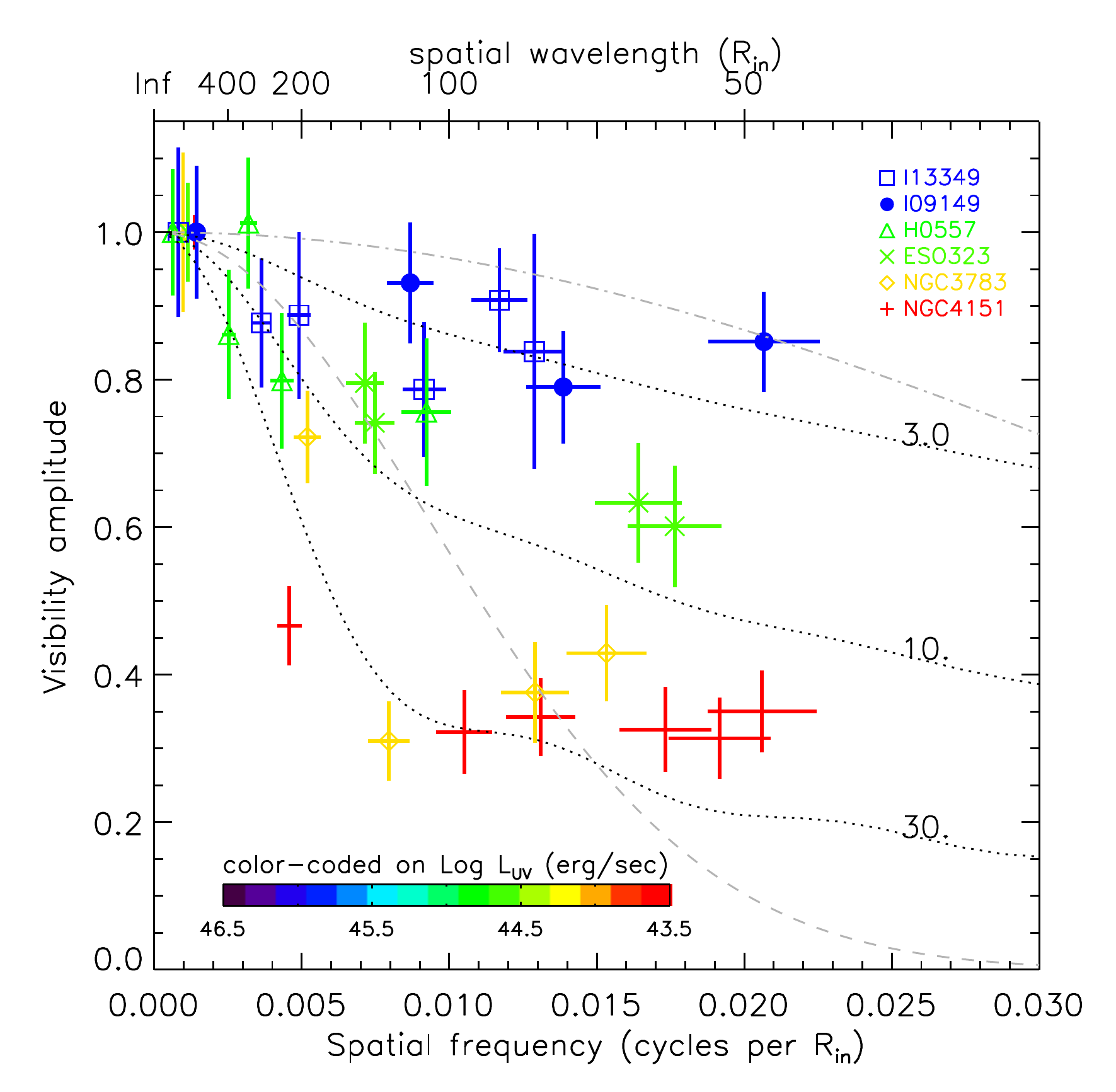}
\caption{\small (a) Observed visibility amplitudes for 6 Type 1 AGNs,
  averaged over 10 to 12 $\mu$m in the rest frame of each object,
  shown as a function of normalized spatial frequency (or its
  reciprocal, normalized spatial wavelength shown in the upper
  axis). The data are plotted using different symbols for different
  objects (see legend), and color-coded using the UV luminosity of the
  central engine in each object (see color bar), labeled with the
  position angle and length of the projected baseline. Dotted curves
  are the visibility functions for power-law brightness distributions
  with designated half-light radii in units of $\Rin$. Dashed and
  dot-dashed gray curves are for Gaussians with HWHM 20 and 5 $\Rin$,
  respectively. (b) The same plot as (a) but with $x$-axis in linear
  scale. Both panels are from \cite{Kishimoto11MIDI}.}
\label{fig_midIR_vis}
\end{figure}

\section{Results: mid-IR}\label{sect-mIR}

Figure~\ref{fig_midIR_vis}$a$ is such a uniform comparison plot. It
shows the visibility amplitudes measured with MIDI/VLTI for 6 Type~1
objects at the rest-frame wavelength range of 10--12 $\mu$m as a
function of normalized spatial frequency in log scale (or normalized
spatial wavelength; see upper axis).  Figure~\ref{fig_midIR_vis}$b$ is
for the reader who prefers to see the same plot in linear scale.

First, {\it if} the structure scales only with $\Lhalf$, i.e. if we
see intrinsically the same distribution (only with the
radius-temperature scaling discussed above), the visibility curves
should look all the same in the normalized units. However, this is
{\it not} the case. Visibilities at a given normalized spatial
frequency, or at a given spatial scale probed, are systematically
different: within the small sample, which nevertheless spans over
$\sim$2.5 orders of magnitudes in UV luminosity, the overall mid-IR
emission size in higher luminosity objects (shown in bluer colors)
seems more compact than that in lower luminosity ones (redder colors),
when seen in units of $\Rin$.

Here we need to be careful when we talk about the size. In
Fig.\ref{fig_midIR_vis}$a$, we also have plotted visibility curves of
Gaussians with HWHM of 20 and 5 $\Rin$. In this log-scale plot, the
visibility curve for a Gaussian with another HWHM will simply be
shifted in the horizontal direction with no change in shape, with the
shift amount proportional to the HWHM ratio.  This conveniently tells
us which visibility data points probe approximately which spatial
scales\footnotemark, but this also immediately tells us
that a single Gaussian seems quite inadequate for describing the
mid-IR visibility curve for each object. This is also evident in
Fig.\ref{fig_midIR_vis}$b$ where none of the objects seem to follow
Gaussian-like visibility curves.

\footnotetext{For instance, the HWHM=20$\Rin$ Gaussian gives
  visibility $V$=0.5 at spatial wavelength
  $\Lambda$=4.5$\times$HWHM=90$\Rin$ (more exactly, 4.5 means
  $\pi/\ln2$), and conversely, an observation at $\Lambda$=90$\Rin$
  would roughly probe a spatial scale of radius $\sim R_{\rm
    V0.5}\equiv\Lambda/4.5$=20$\Rin$.  This quantity $R_{\rm V0.5}$
  would be called a half-visibility radius \cite{Kishimoto11MIDI}.}

Rather, the observed visibility curves show that the mid-IR flux
originates from a much wider range of spatial scales in each
object. The dotted curves in both panels of Fig.\ref{fig_midIR_vis}
are for power-law brightness distributions with different steepnesses,
where the steepness is characterized as a half-light radius, within
which the half of the total integrated light at a given wavelength is
contained (with the inner cut-off at radius $\Rin$; the outer radius
is set to 100$\Rin$ here). Each object seems to follow a power-law
brightness distribution of a different steepness, with higher
luminosity objects looking steeper, while lower luminosity ones being
more extended.  Furthermore, since the radial temperature run of the
heated (illuminated) dust grains as a function of the normalized
radius is expected to be roughly the same for these face-on targets,
the steepness of the brightness distribution would correspond to that
of the surface density of these heated dust grains. Thus we can infer
that the higher luminosity objects have steeper dust density
distributions.

\section{Results: near-IR}\label{sect-nIR}

\begin{figure}
\includegraphics[width=0.6\textwidth]{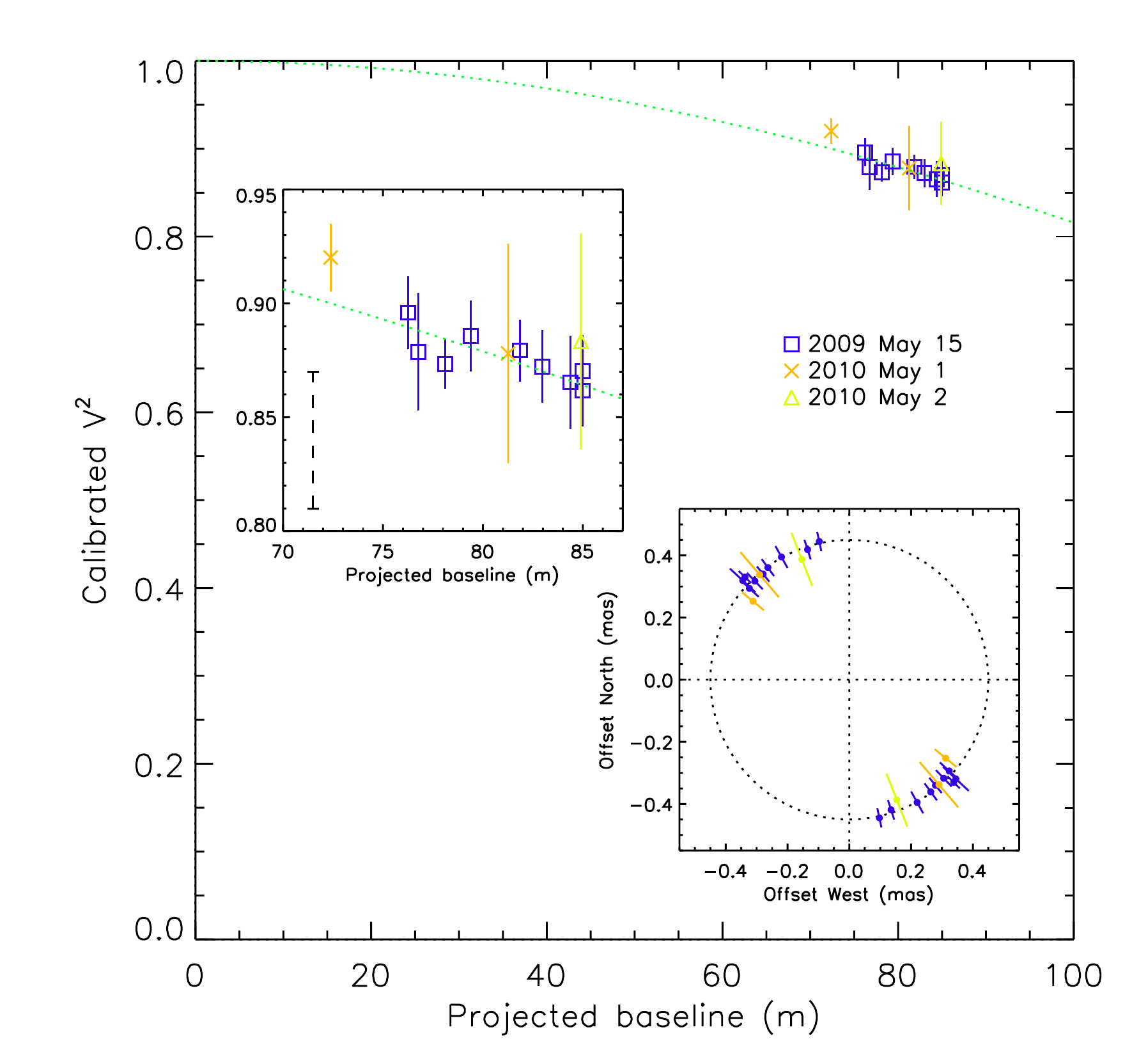}%
\hspace{0.05\textwidth}%
\begin{minipage}[b]{0.35\textwidth}
  \caption{\small Squared visibility $V^2$ of NGC4151 as a function of
    projected baseline lengths observed in two epochs (see legend)
    from \cite{Kishimoto09KI,Kishimoto11}. The possible, perhaps
    conservative, systematic uncertainty in the $V^2$ calibration of
    up to $\sim$0.03 is shown as dashed bar in the left inset (see
    sect.3.2 of \cite{Kishimoto11}). In the right inset, ring radii
    corresponding to each data point are plotted along the position
    angle of each projected baseline. The dotted line in each panel
    corresponds to the best-fit thin-ring model with a radius of 0.45
    mas for our May 2009 data.}
\label{fig_keck_ngc4151}
\end{minipage}
\end{figure}
\begin{figure}
\includegraphics[width=\textwidth]{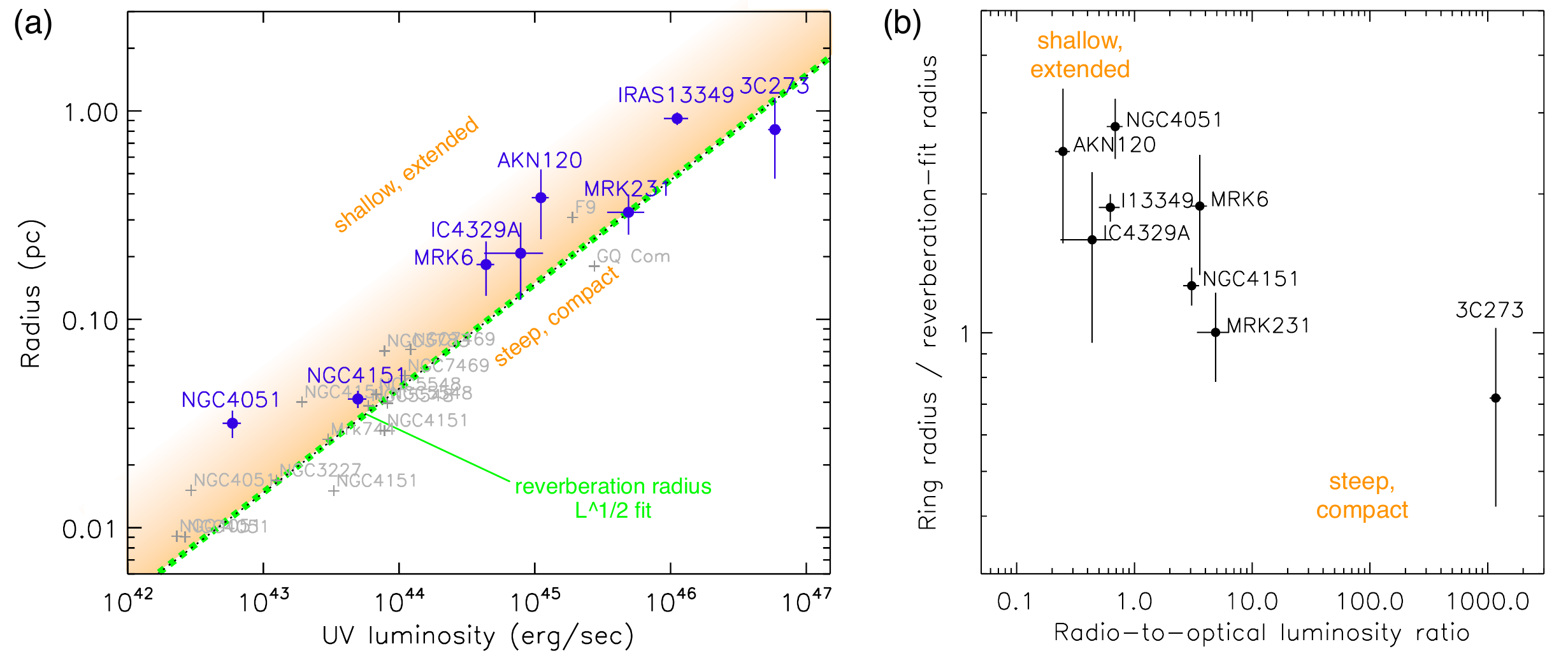}
\caption{\small (a) Interferometric thin-ring radii at 2.2 $\mu$m for
  8 Type 1 AGNs (filled circles \cite{Kishimoto09KI,Kishimoto11}) as a
  function of UV luminosity $L$. The near-IR reverberation radii are
  shown as gray plus signs, and the dotted line is the $L^{1/2}$ fit
  (\cite{Suganuma06}), which defines the dust sublimation radius
  $\Rin$ we adopt here.  The ratio of the ring radius to $\Rin$ would
  indicate the steepness of the brightness distribution, and thus the
  darker and lighter shades correspond to steeper and shallower
  structures, respectively.  (b) The ring radii in units of $\Rin$ are
  plotted against radio (5 GHz) to optical luminosity ratio
  \cite{Kishimoto11}. }
\label{fig_nearIR_vis}
\end{figure}

In the near-IR, the first measurement of the K-band (2.2 $\mu$m)
visibility was made for the brightest Type 1 AGN NGC4151 with the Keck
interferometer (KI), which showed squared visibility $V^2 \sim 0.84$
at a projected baseline $b$ of 83~m \cite{Swain03}. The result was
recently confirmed by us \cite{Kishimoto09KI} and also by
\cite{Pott10}. Furthermore, we measured the K-band visibility for in
total 8 Type 1 AGNs \cite{Kishimoto09KI,Kishimoto11}. All of the
targets showed high visibilities $V^2 \sim$ 0.8 -- 0.9 at $b \sim
80$~m. We interpret this as an indication of the partial resolution of
the dust sublimation region.  First, we have marginally detected a
decrease of visibility over increasing projected baselines for NGC4151
(Fig.\ref{fig_keck_ngc4151}) \cite{Kishimoto09KI,Kishimoto11}, which
would be robust evidence that we are resolving a structure. Secondly,
from this visibility curve, we derived a thin-ring radius for NGC4151,
and also for the other AGNs from the observed visibilities.  These
interferometric ring radii turned out to be quite comparable to the
dust sublimation radii from the near-IR reverberation measurements
(Fig.\ref{fig_nearIR_vis}a) \cite{Kishimoto09KI,Kishimoto11}, which
strongly suggests that we are resolving the dust sublimation region.

A further close look at the ring radii and reverberation radii shows
that the former are slightly and systematically larger than the latter
on average. This is not surprising since the reverberation
measurements are known to be most sensitive to the smallest responding
radii \cite{Koratkar91}, meaning that they are most probably probing a
radius very close to the inner boundary of the dust distribution (thus
adequately adopted as the dust sublimation radius). On the other hand,
the interferometric ring radii are rather brightness-weighted,
effective radii of the overall near-IR brightness
distribution. Therefore, we can postulate that the ratio of the ring
radius to the reverberation radius $\Rin$, i.e. ring radius in units
of $\Rin$, can approximately describe the steepness of the structure,
in the sense that the ratio is closer to unity when the structure is
steeper, while much larger than unity when it is more extended
(Fig.\ref{fig_nearIR_vis}a).

This description of course has a limitation. As the power-law becomes
more shallow and extended, the single ring radius description becomes
more inadequate, as we have seen in the mid-IR visibility
curves\footnotemark.  Here, the underlying assumption is that the
near-IR brightness distribution is steep enough to be approximately
described by a single ring radius, independent of baseline lengths
(though still at relatively low spatial frequencies).  Currently, the
near-IR structure is not well constrained yet, and this should be
followed up with multiple, longer baseline observations.

\footnotetext{Note that we do not differentiate a ring from a
  Gaussian since they have essentially the same visibility curve at
  low spatial frequencies, though a ring model is more physically
  motivated here.}

Under this approximate description of the inner radial structure, do
we see any relationship between the inner radial structure and the
central engine? In fact, there is tentative evidence that
jet-launching objects tend to have a steeper structure
(Fig.\ref{fig_nearIR_vis}$b$) \cite{Kishimoto11}. However, the
statistical significance is not high and this should be investigated
further with a larger sample.

\begin{figure}
\centering
\includegraphics[width=\textwidth]{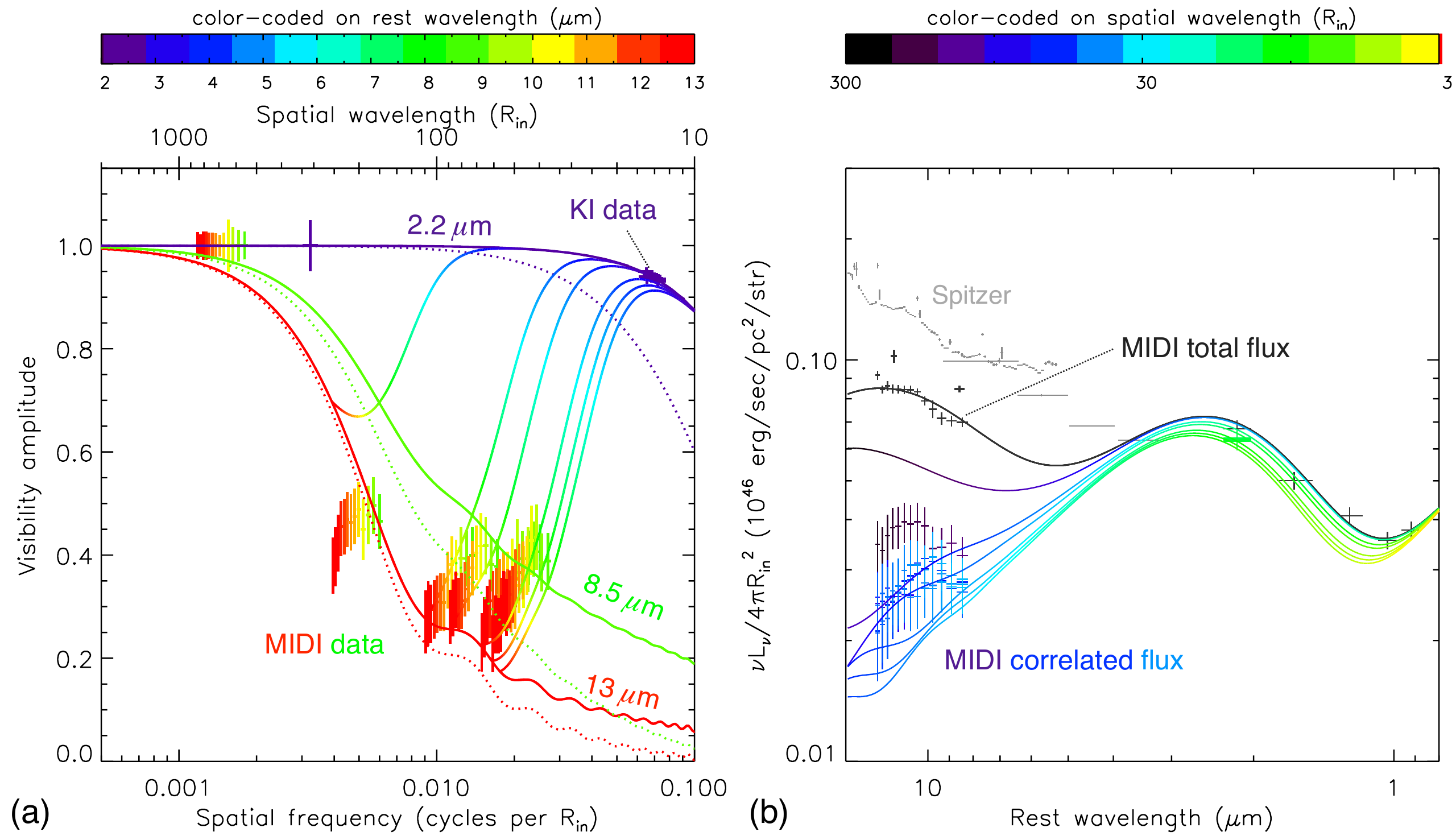}
\caption{\small The mid-/near-IR interferometric and photometric data
  set for NGC4151 \cite{Kishimoto11MIDI}. (a) Observed visibilities as
  color-coded in observing wavelengths as indicated by the color bar
  at the top. The visibilities from a temperature-/density-gradient
  model plus an inner 1400K ring are shown in solid curves for 13,
  8.5, and 2.2 $\mu$m, while dotted curves are the visibilities only
  for the former power-law component.  (b) The total flux spectrum in the
  mid-IR with MIDI and host-galaxy-subtracted total flux spectrum in
  the near-IR are shown in black. The two mid-IR data points slightly
  above the MIDI spectrum are VISIR measurements with a slightly
  larger aperture, and gray data points are the Spitzer measurements
  with even larger apertures. The correlated flux given by the MIDI
  and KI data are shown beneath the total fluxes, color-coded to the
  corresponding spatial wavelength (reciprocal of spatial frequency)
  in units of $\Rin$ (see the color bar at the top). The model curves
  for each MIDI baseline configuration are also shown in the same
  color-coding. See \cite{Kishimoto11MIDI} for more details. }
\label{fig_vis_corrflx}
\end{figure}

\section{Correlated flux as the spectrum of an unresolved structure}

We discussed above the mid-IR and near-IR emission regions separately.
Obviously, we do want to know whether these two are nicely
interconnected.  Can we think these two are coming from a single
component? The answer does not seem to be affirmative, or at least 
not in a very obvious way.

Certainly we do see that the structure looks more and more compact
toward shorter wavelengths. Figure~\ref{fig_vis_corrflx} shows the
data for NGC4151, and some of their aspects are shared by other
objects.  Figure~\ref{fig_vis_corrflx}$a$ shows visibilities both in
the mid- and near-IR, where the color of each data point represents
the observing wavelength as indicated by the color bar at the
top. Toward shorter wavelengths, the overall visibility becomes larger
quite quickly, and thus the corresponding structure becomes smaller.
However, when we look into the total flux spectrum, shown in black in
Figure~\ref{fig_vis_corrflx}$b$, it can be easily seen that if we try
to extrapolate the mid-IR power-law component toward shorter
wavelengths, its color is too red to consistently explain the near-IR
part of the spectrum.  Instead, the near-IR part makes up a somewhat
distinct component on top of the mid-IR power-law
component\footnotemark. As we have seen in the previous section, the
near-IR size scale is quite close to the dust sublimation radius
$\Rin$, and the near-IR total flux spectrum shows that its color
temperature is around 1400~K, quite generically for other objects as
well (e.g.  $\sim$1--2 $\mu$m SED fit in Fig.2 of
\cite{Kishimoto09KI}, and Fig.3 of \cite{Kishimoto07}).  From the
measured size and the absolute flux level of this near-IR component,
its brightness temperature is estimated to be quite close to the color
temperature (see Fig.11 of \cite{Kishimoto11MIDI}), meaning that the
surface filling factor is quite close to unity and the radiation is
almost like that of a black body.  Thus the near-IR flux seems to
originate from an optically-thick, bright rim region of the dust
grains at around $\Rin$ and at the sublimation temperature of
$\sim$1400~K.

\footnotetext{Historically, the bump around 3 $\mu$m in the SED of
  QSOs has been known for decades \cite{Neugebauer79}.}


Figure~\ref{fig_vis_corrflx}$b$ also shows the correlated flux spectra
at progressively longer baseline with different colors (see the color
bar, coded for spatial wavelength).  For the visibilities at
relatively low spatial frequencies in the first lobe (i.e. before the
first null) and in the case of a roughly centro-symmetric object
(which we would expect for Type 1s), the correlated flux approximately
shows the spectrum of the part of the source unresolved by the
interferometer.  The correlated flux at longer baselines excludes the
flux from the outer region, leaving the spectrum only for the inner
core part of the source.  Since we also see that the visibility curves
look relatively flat at long baselines (Fig.\ref{fig_vis_corrflx}$a$),
they probably show the flux fraction of an almost unresolved source,
with the corresponding correlated flux showing the spectrum of this
unresolved core.  The correlated flux spectrum at long baselines in
Fig.\ref{fig_vis_corrflx}$b$ then indicates that the spectrum of this
unresolved core does not seem to be as blue as the mid-IR tail of the
near-IR-emitting hot rim component we inferred above --- it rather
suggests a core at a much lower temperature. In this sense, this lower
temperature core seems to co-exist with the near-IR hot bright rim at
the spatial scale of $\Rin$.

This hybrid description seems also consistent with the mid-IR and
near-IR data for the other 5 targets, having a difference in the
steepness of the mid-IR brightness distribution
\cite{Kishimoto11MIDI}.

\section{Temperature/density gradient parameterization}

All analyses above involve {\it almost no modeling} -- we can
essentially infer all these conclusions from the direct
observables. However, we can also make our arguments a little more
quantitative by introducing a very simple, physical parameterization.
To describe the face-on structure physically, we simply use power-law
distributions of temperature and surface density, as we have already
seen evidence for the distribution resembling a power law. In
addition, we approximate the near-IR rim emission as a ring with a
certain effective radius. The accretion disk component, which seems to
become important only at wavelengths $\lesssim$ 1~$\mu$m, is assumed
to be unresolved and have the near-IR spectral shape of $f_{\nu}
\propto \nu^{+1/3}$, which is supported both theoretically and
observationally \cite{Koratkar99,Kishimoto08}.

\begin{figure}
\centering
\includegraphics[width=1.0\textwidth]{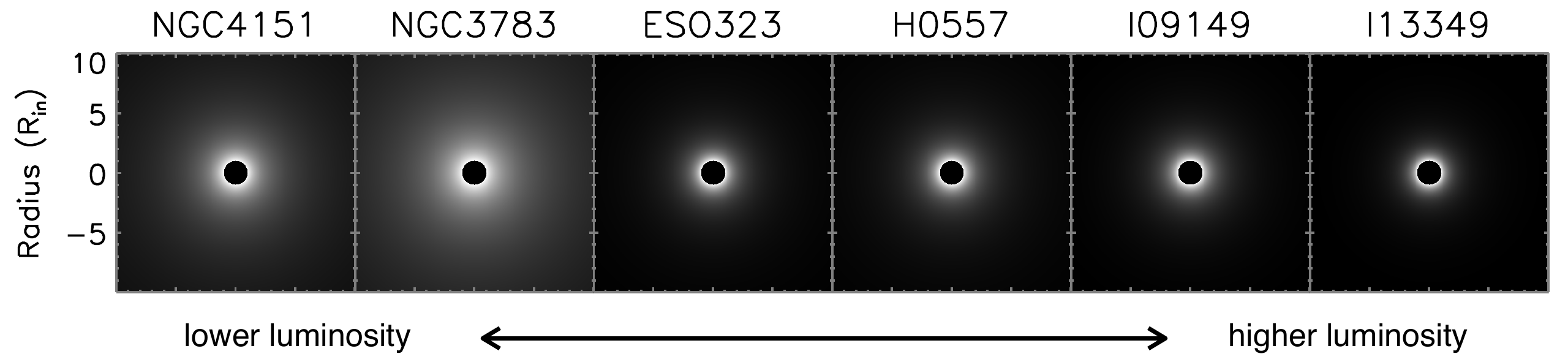}
\caption{\small Model images of the power-law component at 13 $\mu$m
  in the rest frame of each object, shown in linear scale, using the best-fit parameters for
  the temperature and surface density indices \cite{Kishimoto11MIDI},
  with spatial scales shown in units of $\Rin$. Higher luminosity
  objects are put toward the right.}
\label{fig_image_rin}
\end{figure}


\begin{figure}
\includegraphics[width=0.6\textwidth]{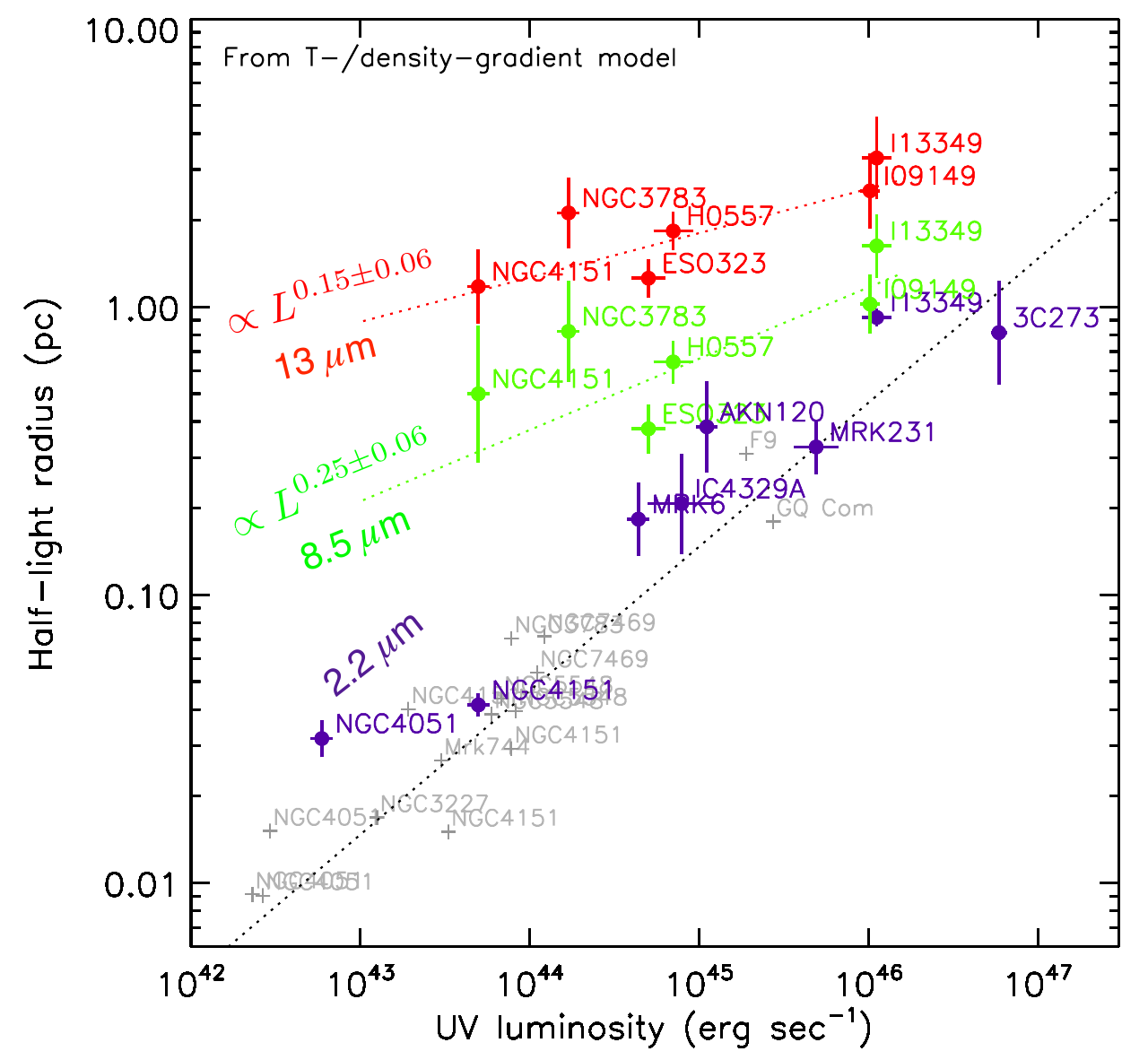}%
\hspace{0.05\textwidth}%
\begin{minipage}[b]{0.35\textwidth}
  \caption{\small The mid-IR half-light radii $\Rhalf$ in pc from
    best-fit models at rest-frame 13 and 8.5 $\mu$m (red and green,
    respectively), plotted against UV luminosity.  Thin-ring radii in
    pc at 2.2 $\mu$m are plotted in purple for the KI-observed
    sample. The near-IR reverberation radii (\cite{Suganuma06} and
    references therein) are indicated as gray plus signs with the
    $\Lhalf$ fit shown as a black dotted line (which is our definition
    of $\Rin$).  The mid-IR part of this Figure is more
    model-dependent than, but essentially consistent with, that of
    Fig.8 in \cite{Kishimoto11MIDI}. The fits to $\Rhalf$ are shown in
    red and green dotted lines with quoted slopes, where the
    normalizations are 2.6$\pm$0.5 and 1.2$\pm$0.2 pc at $L\equiv 6
    \nu L_{\nu}$(5500\AA)=10$^{46}$ erg/sec for 13 and 8.5 $\mu$m,
    respectively.  }
\label{fig_model}
\end{minipage}
\end{figure}

This parameterization can describe the whole data sets at least
approximately.  Figure~\ref{fig_vis_corrflx} includes the fitted model
curves for NGC4151, and those for the whole sample are found in Figs.1
and 2 of \cite{Kishimoto11MIDI}.  Using the best-fit parameters, we
can construct the model images which visually illustrate what we have
discussed.  The panels of Figure~\ref{fig_image_rin} are such images
at 13 $\mu$m in the rest frame of each object, with spatial scales in
units of $\Rin$, showing that the structure becomes steeper for higher
luminosity objects (to the right).  We can also show this structural
change back in unnormalized, physical spatial scales, e.g. in pc, by
looking at the change of characteristic radii at different
wavelengths.  Figure~\ref{fig_model} shows the mid-IR half-light radii
in pc for the 6 mid-IR targets at the rest-frame 13 and 8.5~$\mu$m
derived from the model fits above, together with the thin-ring radii
at 2.2~$\mu$m for the 8 near-IR targets (with two overlapping
targets).  If the surface density structure were the same over the
sample, the ratios of the mid-IR to near-IR size would just be the
same over the sample.  However, since mid-IR radii in pc increase with
luminosity~$L$ {\it much slower} than $\Lhalf$ (specific values are
given in Fig.\ref{fig_model}), the difference between radii at
different wavelengths becomes smaller toward higher luminosities,
meaning that the radial density structure becomes steeper at higher
luminosities.

Note that Figure~\ref{fig_model} is essentially the same as Fig.8 of
\cite{Kishimoto11MIDI} which shows the mid-IR half-light radii measured
with a simple geometrical power-law brightness at each wavelength. The
half-light radii from the temperature-/density-gradient model are
certainly more model-dependent, but utilize more information for each
object including the total flux spectrum. The results are essentially consistent
within the errors.


The structural steepening could potentially mean that, since the
accretion disk luminosity is directly linked to the mass accretion
rate, the radial density distribution might become steeper in higher
accretion rate objects (here we refer to the absolute rate, not
normalized by the Eddington rate).  We now even have quantitative
estimations for the radial surface density index of the heated dust
distribution, which ranges from $\sim$$r^{0}$ to $\sim$$r^{-1}$ for
the luminosities covered.  We plan to further investigate correlations
with different quantities such as the Eddington ratio using a larger
sample in order to advance our physical understanding.

\section{Conclusions}

We have summarized the results of our first systematic study of Type 1
AGNs with IR interferometry, where we attempted to map the radial
structure of the inner dusty accreting material. For the warm mid-IR
power-law-like component, we seem to see steeper density distributions
in objects with higher luminosities, or higher mass accretion rates.
For the near-IR compact component in the dust rim region, we see a
tentative correlation of steeper structures with more radio-loud
central engines.  We plan to further scrutinize all the results with a
larger sample.


\begin{thebibliography}{10}
\expandafter\ifx\csname url\endcsname\relax
  \def\url#1{{\tt #1}}\fi
\expandafter\ifx\csname urlprefix\endcsname\relax\def\urlprefix{URL }\fi
\providecommand{\eprint}[2][]{\url{#2}}

\bibitem{Suganuma06}
{Suganuma} M, {Yoshii} Y, {Kobayashi} Y, {Minezaki} T, {Enya} K, {Tomita} H,
  {Aoki} T, {Koshida} S and {Peterson} B~A 2006 {\em \apj\/} {\bf 639} 46--63
  (\textit{Preprint} \eprint{arXiv:astro-ph/0511697})

\bibitem{Kishimoto09}
{Kishimoto} M, {H{\"o}nig} S~F, {Tristram} K~R~W and {Weigelt} G 2009 {\em
  \aap\/} {\bf 493} L57--L60 (\textit{Preprint} \eprint{0812.1964})

\bibitem{Kishimoto07}
{Kishimoto} M, {H{\"o}nig} S~F, {Beckert} T and {Weigelt} G 2007 {\em \aap\/}
  {\bf 476} 713--721 (\textit{Preprint} \eprint{arXiv:0709.0431})

\bibitem{Barvainis87}
{Barvainis} R 1987 {\em \apj\/} {\bf 320} 537--544

\bibitem{Maiolino01II}
{Maiolino} R, {Marconi} A and {Oliva} E 2001 {\em \aap\/} {\bf 365} 37--48
  (\textit{Preprint} \eprint{arXiv:astro-ph/0010066})

\bibitem{Gaskell04}
{Gaskell} C~M, {Goosmann} R~W, {Antonucci} R~R~J and {Whysong} D~H 2004 {\em
  \apj\/} {\bf 616} 147--156 (\textit{Preprint}
  \eprint{arXiv:astro-ph/0309595})

\bibitem{Kawaguchi10}
{Kawaguchi} T and {Mori} M 2010 {\em \apjl\/} {\bf 724} L183--L187
  (\textit{Preprint} \eprint{1010.5799})

\bibitem{Kishimoto11MIDI}
{Kishimoto} M, {H{\"o}nig} S~F, {Antonucci} R, {Millour} F, {Tristram} K~R~W
  and {Weigelt} G 2011 {\em \aap\/} {\bf 536} A78 (\textit{Preprint}
  \eprint{1110.4290})

\bibitem{Swain03}
{Swain} M, {Vasisht} G, {Akeson} R, {Monnier} J, {Millan-Gabet} R, {Serabyn} E,
  {Creech-Eakman} M, {van Belle} G, {Beletic} J, {Beichman} C, {Boden} A,
  {Booth} A, {Colavita} M, {Gathright} J, {Hrynevych} M, {Koresko} C, {Le
  Mignant} D, {Ligon} R, {Mennesson} B, {Neyman} C, {Sargent} A, {Shao} M,
  {Thompson} R, {Unwin} S and {Wizinowich} P 2003 {\em \apjl\/} {\bf 596}
  L163--L166 (\textit{Preprint} \eprint{arXiv:astro-ph/0308513})

\bibitem{Kishimoto09KI}
{Kishimoto} M, {H{\"o}nig} S~F, {Antonucci} R, {Kotani} T, {Barvainis} R,
  {Tristram} K~R~W and {Weigelt} G 2009 {\em \aap\/} {\bf 507} L57--60
  (\textit{Preprint} \eprint{0911.0666})

\bibitem{Pott10}
{Pott} J, {Malkan} M~A, {Elitzur} M, {Ghez} A~M, {Herbst} T~M, {Sch{\"o}del} R
  and {Woillez} J 2010 {\em \apj\/} {\bf 715} 736--742 (\textit{Preprint}
  \eprint{1003.4757})

\bibitem{Kishimoto11}
{Kishimoto} M, {H{\"o}nig} S~F, {Antonucci} R, {Barvainis} R, {Kotani} T,
  {Tristram} K~R~W, {Weigelt} G and {Levin} K 2011 {\em \aap\/} {\bf 527} A121
  (\textit{Preprint} \eprint{1012.5359})

\bibitem{Koratkar91}
{Koratkar} A~P and {Gaskell} C~M 1991 {\em \apjs\/} {\bf 75} 719--750

\bibitem{Neugebauer79}
{Neugebauer} G, {Oke} J~B, {Becklin} E~E and {Matthews} K 1979 {\em \apj\/}
  {\bf 230} 79--94

\bibitem{Koratkar99}
{Koratkar} A and {Blaes} O 1999 {\em \pasp\/} {\bf 111} 1--30

\bibitem{Kishimoto08}
{Kishimoto} M, {Antonucci} R, {Blaes} O, {Lawrence} A, {Boisson} C, {Albrecht}
  M and {Leipski} C 2008 {\em \nat\/} {\bf 454} 492--494 (\textit{Preprint}
  \eprint{arXiv:0807.3703})

\end{thebibliography}

\providecommand{\newblock}{}

\end{document}